# Wenzhou TE: a first-principles calculated thermoelectric materials database

Ying Fang [1] and Hezhu Shao [1*]

**Abstract:** Since the implementation of the Materials Genome Project by the Obama administration in the United States, the development of various computational materials databases has fundamentally expanded the choices of industries such as materials and energy. In the field of thermoelectric materials, the thermoelectric figure of merit ZT quantifies the performance of the material. From the viewpoint of calculations for vast materials, the ZT values are not easily obtained due to their computational complexity. Here, we show how to build a database of thermoelectric materials based on first-principles calculations for the electronic and heat transport of materials. Firstly, the initial structures are classified according to the values of bandgap and other basic properties using the clustering algorithm K-means in machine learning, and high-throughput first principles calculations are carried out for narrow-bandgap semiconductors which exhibiting potential thermoelectric application. The present framework of calculations mainly includes deformation potential module, electrical transport performance module, mechanical and thermodynamic properties module. We have also set up a search webpage for the calculated database of thermoelectric materials, providing searching and viewing the related physical properties of materials. Our work may inspire the construction of more computational databases of first-principle thermoelectric materials and accelerate research progress in the field of thermoelectrics.

School of Electrical and Electronic Engineering, Wenzhou University, Zhejiang, 325035, China. (hzshao@wzu.edu.cn)

# 1. Introduction

In 2011, the Obama administration of the United States officially proposed the "Material Genome Project", which utilizes high-throughput computing and experiments to obtain massive material data, combined with data analysis technology by artificial intelligence for new material development. The goal is to shorten the cycle of new materials development and applications, as well as reduce the costs for materials research and development, so that the United States can continue to maintain a leading position in manufacturing technology. In 2016, the US government released the "First Five Years of the Materials Genome Initiative: Accommodations and Technical Highlights" report, which pointed out that during the five years of the implementation of the Materials Genome Engineering program, federal research institutions such as the Department of Energy, the Department of Defense, the Natural Science Foundation, the National Bureau of Standards and Technology, and the National Aeronautics and Space Administration have invested over 500 million US dollars, establishing computational materials research and development centers including the National Network for Virtual High throughput Preparation (NIST&NREL) and the Center for Cross scale Material Design and Multi scale Materials Research (NIST, ANL, ARL), forming three major computational materials databases: the Materials Project (MP) [1], AFLOW [2], and OQMD [3,4], several auxiliary databases such as Materials Data Repository (MDR), Materials Resource Registry, Energy Materials Network, as well as databases related analysis tools.

Shortly after the proposal of the Materials Genome Project by the United States, the European Science Foundation launched the Accelerated Metallurgy (ACCMET) program, which costs over 2 billion euros, with the aim of keeping up with the pace of the United States. The European Commission funded the Horizon 2020 project NoMatD, led by the Max Planck Institute in German, for a period of three years in 2015. The project aims to use the "centralized data warehouse" method to involve various research groups and provide data related to computational materials science, with the aim of building a "Encyclopedia of Materials" and a tool for analyzing big data on materials. In the UK, the government has also implemented the e-science program, with its funding, to carry out high-throughput material computing simulations and the construction of material computing basic databases, such as eMinerals and the "Material Grid" project. The Swiss EPFL University has led the development of the European Materials Database AiiDA [5].

Nowadays, with the vigorous development of big data and artificial intelligence technology, the material genome project research characterized by high-throughput experiments, high-throughput computing, and artificial intelligence big data analysis is in full swing, and has shown astonishing advantages in many materials fields. The paper "Machine-learning-assisted materials discovery using failed experiments" published in Nature in May 2016 [6] showed that based on years of accumulated experimental data, various catalytic new materials can be discovered using artificial intelligence (AI) technology. This work indicates that AI will profoundly transform the research methods in the field of materials. The centuries long history of human scientific development has formed three research paradigms: experimental, theoretical, and computational. However, in the fields of complex systems such as biology, astronomy, and materials, there are very complex interactions involved, coupled with a large number of variables, which greatly limits the effectiveness of theoretical and computational research models and requires the combination of big data and AI as the "fourth paradigm". In 2017, AlfaGo defeated the human Go master, but Google disbanded the DeepMind team responsible for developing the program, and then formed an AI research and development team engaged in material genome engineering. At present, American high-tech companies including Apple, Google, IBM, Tesla, etc. are all laying out the use of AI for the research and development of new materials based on material genomics methods. The fourth paradigm of materials science requires the ability to generate and process massive amounts of data, thus obtaining massive amounts of material data has become a key aspect of the Materials Genome Project. With the improvement of computing power, the accumulation of material data based on high-throughput computing is receiving more and more attention, and its application in the research and development of new thermoelectric materials is expected to greatly accelerate its application process.

The performance of thermoelectric materials is described by the figure of merit ZT, which can be expressed as follows:

$$ZT = \frac{S^2 \sigma T}{\kappa_e + \kappa_l} \qquad (1)$$

Where $S$ is the Seebeck coefficient, $\sigma$ is the conductivity, $T$ is the temperature, $\kappa_e$ and $\kappa_l$ is the thermal conductivity contributed by carriers and phonons, respectively. These parameters of $S$, $\sigma$ and $\kappa$ are coupled with each other, and it is difficult to independently regulate them. For example, for semiconductor materials, increasing doping concentration can increase conductivity, while at the same time reducing the Seebeck coefficient and increasing carrier thermal conductivity. At present, the three major material databases, Materials Project, AFLOW, and OQMD, have data on several common physical quantities, including atomic and band structure, and other physical properties are also being added. However, thermoelectric performance of materials, due to their particularity and the complexity in calculating electrical and thermal transport properties, generally require a large amount of computation.

Here we selects Materials Project as the structural source for constructing a thermoelectric material database. Specifically, we employed the atomic structure files POSCAR and CIF (currently 19952 materials) in MP materials with id-number below 100000 through the Materials Project API as the initial materials for building present thermoelectric material database——**Wenzhou TE**. We have built deformation potential modules, elastic properties modules, and BoltzTrap electronic transport modules. And then, we collect data by Python scripts and display it on a web site, https://hezhu2024.github.io, for others to use.

**2. Methodology**

2.1. Clustering (K-means)

At present, the excellent thermoelectric materials obtained in experiments are mainly semiconductors with narrow-bandgaps, then we choose bandgap as a major feature for material screening. At the same time, we selected free energy, volume, density, and average atomic energy as other features from the descriptors obtained from the MP database. They form five featured variables for the K-means clustering algorithm.

Here is a brief introduction to the K-means principle [7]. K-means is a clustering algorithm that divides data into K classes. Firstly, K class random points are randomly generated, denoted as $O_1, O_2, \cdots O_l, \cdots O_K$. Assuming that the *j*-th feature of the *i*-th data is represented as $x_{ij}$, the distance from the *i*-th data sample to the *l*-th class random point is:

$$d_{il} = \sqrt{\sum_{j=0}^{j=J}(x_{ij} - O_{lj})^2} \qquad (2)$$

Among them, *J* represents a total of *J* features in the data. The random class point with the smallest distance represents the same class. After the first iteration, each data sample will be classified into a certain class. Then, we calculate the average value of each class of data as the new random class point. The new random class point can be represented as:

$$O_{lj} = \frac{1}{N}\sum_{i=0}^{i=N} x_{ij} \qquad (3)$$

among them, $j \in [1,2,\ldots,J], l \in [1,2,\ldots,K]$.

Then we re-calculate these distances, and reclassify them. And such process is repeated until convergence achieved. And finally the data will be classified into K classes. In present work, we also standardize the data before classification. In order to illustrate how many categories are most reasonable, we could assume that the formula for the total loss as follows:

$$Loss = \sum_{i=0}^{i=n} d_{il} \qquad (4)$$

Where n represents the number of samples. This formula represents the sum of distances from all sample points to their random class points. When there is a significant inflection point on the line of Loss with respect to class K, the value of K at the inflection point should be considered as a reasonable classification. Through the K-means method, we divided the initial materials from MP into 5 categories. Their quantities are 6602, 5425, 3770, 2800, and 1355, respectively.

*2.2. Deformation Potential Theory (DPT)*

The deformation potential theory was proposed by Bardeen and Shockley [8] in the 1950s to describe charge transfer in non-polar semiconductors. The charge mobility can be expressed as $\mu_x = e\tau_x/m^*$, where the relaxation time for bulk materials could be written as follows [8,9]

$$\tau_x = \frac{2\sqrt{2\pi}\hbar^4 C_x}{3(k_B T m^*)^{\frac{3}{2}} E_{DPx}^2} \qquad (5)$$

where $C_x = \partial^2 E/(\partial(\Delta a_x/a_x)^2 V_0)$ is the elastic constant, $E_{DPx} = \Delta V_i/(\Delta a_x/a_x)$, $\Delta V_i$ is the deformation potential energy, which is the difference between the energy level of the *i*-th energy band and the energy level of the deep nuclear state, and $m^* = \hbar^2/(\partial^2 E/\partial k^2)$ is the effective mass.

## 2.3. Elastic and thermal properties

We can obtain elastic properties, group velocity, Poisson's ratio, Debye temperature, Grüneisen coefficients, and lattice thermal conductivity, by after calculating the elastic constant of materials [10], which could be easily achieved for the high-throughput calculation.

In the case of uniform deformation for a crystal, the generalized form of Hooke's law of stress-strain [11] is:

$$f_{ij} = C_{ijkl}\epsilon_{kl} \qquad (6)$$

where $f_{ij}$ and $\epsilon_{kl}$ is a homogeneous second-order stress tensor and a strain tensor, respectively [12]. $C_{ijkl}$ represents the fourth order elastic stiffness tensor. Using matrix representation, we can abbreviate the stiffness tensor $C_{ijkl}$ of four suffixes to the stiffness tensor $C_{ij}$ of two suffixes, which can be represented as follows:

$$C_{ij} = \begin{bmatrix} C_{11} & C_{12} & C_{13} & C_{14} & C_{15} & C_{16} \\ C_{21} & C_{22} & C_{23} & C_{24} & C_{25} & C_{26} \\ C_{31} & C_{32} & C_{33} & C_{34} & C_{35} & C_{36} \\ C_{41} & C_{42} & C_{43} & C_{44} & C_{45} & C_{46} \\ C_{51} & C_{52} & C_{53} & C_{54} & C_{55} & C_{56} \\ C_{61} & C_{62} & C_{63} & C_{64} & C_{65} & C_{66} \end{bmatrix} \qquad (7)$$

Similarly, the elastic flexibility tensor $(s_{ij} = C_{ij}^{-1})$ can be written as:

$$s_{ij} = \begin{bmatrix} s_{11} & s_{12} & s_{13} & s_{14} & s_{15} & s_{16} \\ s_{21} & s_{22} & s_{23} & s_{24} & s_{25} & s_{26} \\ s_{31} & s_{32} & s_{33} & s_{34} & s_{35} & s_{36} \\ s_{41} & s_{42} & s_{43} & s_{44} & s_{45} & s_{46} \\ s_{51} & s_{52} & s_{53} & s_{54} & s_{55} & s_{56} \\ s_{61} & s_{62} & s_{63} & s_{64} & s_{65} & s_{66} \end{bmatrix} \qquad (8)$$

The Voigt [13] Bulk modules can be calculated by

$$B_v = \frac{1}{9}[(C_{11} + C_{22} + C_{33}) + 2(C_{12} + C_{23} + C_{31})] \qquad (9)$$

And the Shear modulus can be obtained by

$$G_v = \frac{1}{15}[(C_{11} + C_{22} + C_{33}) - (C_{12} + C_{23} + C_{31}) + 3(C_{44} + C_{55} + C_{66})] \qquad (10)$$

The Reuss [14] Bulk and Shear modulus can be calculated by

$$\frac{1}{B_r} = (s_{11} + s_{22} + s_{33}) + 2(s_{12} + s_{23} + s_{31}) \qquad (11)$$

and

$$\frac{15}{G_r} = 4(s_{11} + s_{22} + s_{33}) - 4(s_{12} + s_{23} + s_{31}) + 3(s_{44} + s_{55} + s_{66}) \qquad (12)$$

In present work, we take the arithmetic mean of the boundaries between Voigt and Reuss Voigt-Reuss-Hill (VRH) [15]:

$$B_h = \frac{B_r + B_v}{2} \qquad (13)$$

$$G_h = \frac{G_r + G_v}{2} \qquad (14)$$

The longitudinal ($v\_l$), transverse ($v\_t$), and average ($v\_a$) elastic wave velocities can be calculated by

$$v_l = \sqrt{\frac{3B_h + 4G_h}{3\rho}}, \quad v_t = \sqrt{\frac{G_h}{\rho}}, \quad v_a = \left[\frac{1}{3}\left(\frac{2}{v_t^3} + \frac{1}{v_l^3}\right)\right]^{-1/3} \qquad (15\sim17)$$

The Debye temperature ($\theta_D$) is obtained by:

$$\theta_D = \frac{h}{k_B}\left[\frac{3q}{4\pi}\frac{N\rho}{M}\right]^{1/3} v_a \qquad (18)$$

And the Grüneisen coefficient is calculated by:

$$\gamma = \frac{3}{2}\left(\frac{1 + v_{poi}}{2 - 3v_{poi}}\right) \qquad (19)$$

Where $v_{poi} = (1 - 2\left(\frac{v_t}{v_l}\right)^2)/(2 - 2\left(\frac{v_t}{v_l}\right)^2)$ is the Poisson's ratio.

According to the Slack formula [16,17], the lattice thermal conductivity can be expressed as:

$$k_l = A\frac{\bar{M}\theta_D^3 \delta}{\gamma^2 n^{2/3} T} \qquad (20)$$

where $\bar{M}$ is the average atomic mass, $\theta_D$ is the Debye temperature, $\delta$ is the volume of each atom, $n$ is the number of atoms in the original cell, $\gamma$ is the Grüneisen coefficient, $A$ is a constant of $3.1 \times 10^{-6}$, and $T$ is the temperature.

### 2.4. Methods for the first-principles calculations and transport properties

In the process of building a thermoelectric material database, first-principles calculations are done by the Vienna Ab initio Simulation Package (VASP)[18,19]. The calculation of electricity transportation requires the use of the Boltztrap program package [20]. In order to minimize computational costs while ensuring data reliability, during optimizing calculations, we set the plane-wave energy cutoff to be 1.4 times the maximum ENMAX of POTCAR of composed elements, the electronic energy convergence to be $10^{-4}$ eV, the force convergence for ions to be $10^{-2}$ eV/Å, and the density k-mesh to be 0.04×2π Å$^{-1}$.

All the processed are controlled through Shell scripts. Data collection and calculation are implemented by Python scripts. These codes are home-made.

## 3. Capabilities and workflow

### 3.1. The application of K-means on datasets from MP

From Figure 1a, it can be seen that the number of points with obvious inflection is 6, which means that the initial structures can be divided into 6 categories. Considering the reasonable distribution of the average-bandgap values, we ultimately divided it into 5 categories. The featured distribution map and various information of K-means are shown in Figures 1c-g. The average value of bandgap for the first class is merely 0.025eV, so this class of material contains many metals. The second class with average bandgap value of 0.14eV mainly composed of semiconductors with narrow bandgaps. The third, fourth, and fifth categories are mainly composed of semiconductors and insulators with wide bandgaps. As a starting point, we focused on calculating the physical properties of candidate material sets for the first and second categories.

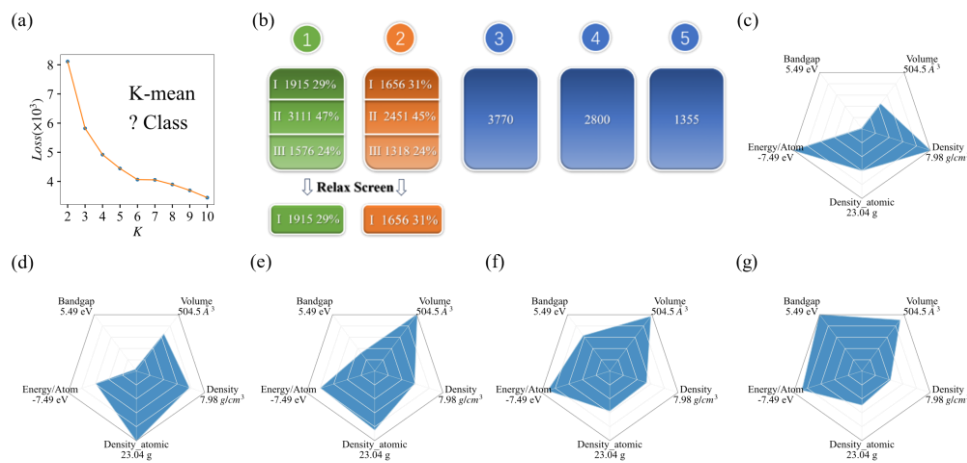

**Figure 1.** The application of K-means on MP databases: (a) The line chart of Loss for K-class; (b) the classification data and relaxation screening results of the initial structures under K-means; (c-g) the average distribution of 5 features for each K-means class.

*3.2. Computational framework and relaxation process*

After getting the structural file, we firstly perform structural relaxation and static calculation. Structural relaxation refers to the optimization process of atomic positions and lattice constants. We employed VASP software for the first-principles calculations. Actually several mainstream databases such as AFLOW, MP, OQMD, etc. are also calculated using VASP software.

For the first and second types of materials obtained through K-means initial screening, there are more than 12000 materials, many of which contain too many element types and numbers of atoms in the primitive cell. In present work, we firstly calculate the material system with a relatively simple structure. Therefore, a computational control process is employed during the structural relaxation to further screen them, and resulting in a total of more than 3000 materials with relatively simple structures in the first and second types. Nevertheless, conducting structural relaxation for so many materials is a computationally demanding task. In order to accelerate the calculation, we wrote several shell scripts to control the process of structural relaxation. The flowchart is shown in Figure 2.

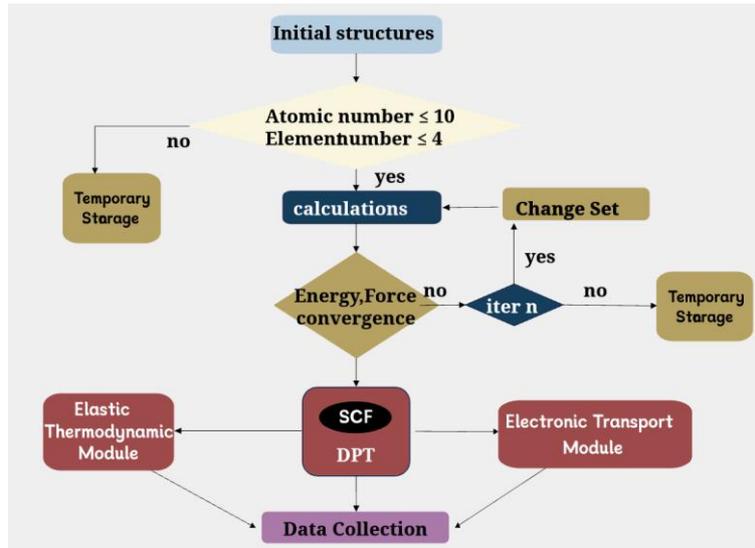

**Figure 2.** Flowchart for constructing thermoelectric material database.

After performing relaxation calculations on the data of the first and second classes of materials, we screened 1915 and 1656 materials, respectively, for further calculations, as shown in Figure 1b. In the first class, there are remained 3111 materials with atomic numbers greater than 10 or element types greater than 4, and other 1576 materials unrelaxed structures which are hard to get convergent relaxation in our present setup calculations. In the second category, there are also 2451 materials with atomic numbers greater than 10 or element types greater than 4, and 1318 materials that are difficult to be relaxed. After the relaxation calculation process, the convergent structures are saved for further calculations.

Then we perform the calculations of the parameters of deformation potential theory. Firstly, we performed an anisotropic property judgment on the material, and then we performed static calculations on the deformed structures in various directions.

3.3. *Analysis of results of deformation potential theory (using Si as an example)*

The deformation potential method considered acoustic phonons as the main scattering sources for electrons. The relaxation time obtained by ignoring the contributions of optical phonon branches and other scattering mechanisms could be larger than the real one, but the calculation of deformation potential is relatively simple, easily employed in high-throughput calculations. The coefficients for applying deformation to the lattice vector are {0.98, 0.99, 1.00, 1.01, 1.02} of relaxed volumes, respectively. Such calculations could ensure the reliability of fitting with the second-order function for the elastic constant and the first-order function for the elastic potential energy. Taking Si as an example, as shown in Figure 3.

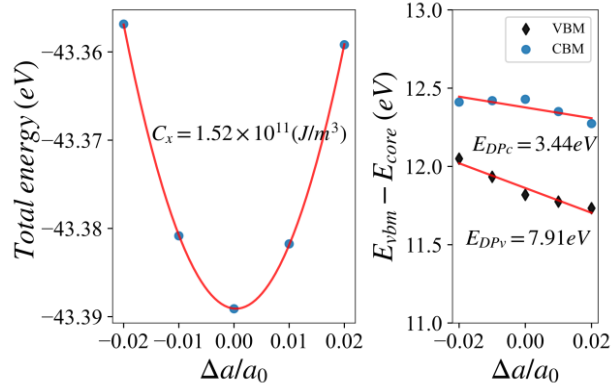

**Figure 3.** Schematic diagram of second-order fitting elastic constant $C_x$ and first-order fitting of elastic potential energy $E_{DP}$ for Si.

After calculating the deformation potential parameters, we could get the relaxation time of carriers by combing the effective masses.

**Table 1.** Calculated deformation potential parameters, effective mases and relaxation time of carries for Si.

|  | Carrier | $E_{DPx}$ (eV) | $C_x(10^{11} \text{Jm}^{-3})$ | $m^*/m_0$ | $\tau_x$(fs) |
|---|---|---|---|---|---|
| Si | Electron | 3.44 | 1.52 | 0.46 | 1141.9 |
|  | Hole | 7.91 | 1.52 | 2.48 | 21.6 |

*3.4. Energy band and effective mass calculation*

There are many methods to obtain the band structure of a material. Here we compare three feasible schemes. The first scheme is VASP high symmetry point energy band calculation, the second one is using BoltzTrap2 [20] to fit the band structure, and the third one is using maximally-localized Wannier function to interpolate the VASP results [21]. Considering the accuracy and efficiency, the second scheme is chosen in our high-throughput calculations. As shown in Table 2, three schemes for Si are presented.

**Table 2.** Band results of Si under three schemes.

|  | VASP | Boltztrap | Wannier90 |
|---|---|---|---|
| Bandgap (eV) | 0.61 | 0.59 | 0.71 |
| $m_c^{*1}/m_0$ | 0.97 | 0.46 | 0.55 |
| $m_v^{*2}/m_0$ | 2.63 | 2.48 | 2.03 |
| **User time (s)** | **8.048** | **1.057** | **6.103** |
| Cores of Cpu | 10 | 10 | 10 |

[1] $m_c^*$ is effective mass of conduction band.

[2] $m_v^*$ is effective mass of price band.

The bandgap of Si in the MP database is 0.61eV, which is consistent with VASP calculation. The bandgap error calculated by Boltztrap is within 5%. Meanwhile, the effective mass of Si calculated by Boltztrap is smaller than that of the VASP scheme, indicating that the calculated relaxation time will be larger, as shown in Table 1, where the relaxation time of electrons is $1141.9fs$. The energy band of Si by three schemes is shown in Figure 4. From Table 2, it can be seen that the Boltztrap calculation for band structure is most efficient, then it can help to accelerate the high-throughput calculation.

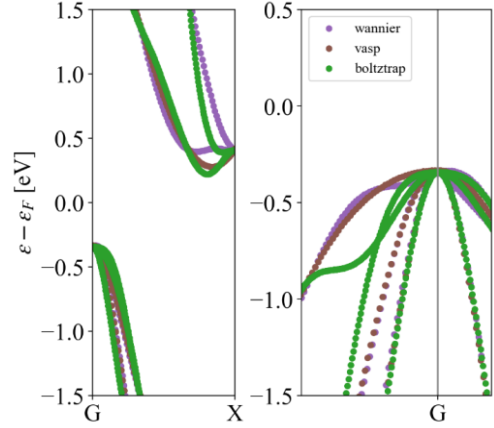

**Figure 4.** The band structure of Si by three schemes: VASP, Boltztrap, and Wannier interpolations.

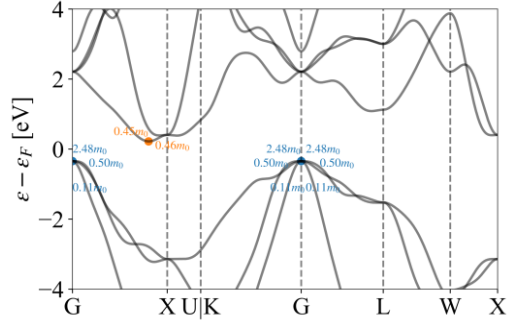

**Figure 5.** The effective mass of Si by the Boltztrap scheme.

To facilitate high-throughput calculation, we use the formula $m^* = \hbar^2/(\partial^2 E/\partial k^2)$ to calculate the effective mass. The effective masses of Si by the Boltztrap scheme is shown in Figure 5. A series of effective masses of conduction and valence bands were obtained near the high symmetry points of Γ and X. We selected the maximum values of $0.46m_0$ and $2.48m_0$ as the effective masses for the conduction band and valence band, respectively. In addition, our program is designed to automatically determine whether the band is degenerate and calculate the effective mass for each degenerate band. We note here that the reason for selecting the maximum effective mass is that the deformation potential overestimates the relaxation time. By selecting the maximum effective mass, the relaxation time can be effectively reduced to compensate for the shortcomings of the deformation potential theory. In high-throughput calculations, the program also selects representative effective masses for other materials such as the Si.

*3.5. High-throughput electrical transport properties(Boltztrap)*

Boltztrap is a program package calculating the semi-classic transport coefficients, based on a smoothed Fourier interpolation of the bands. Electrical transport properties such as Seebeck coefficient, electronic conductivity, and electronic thermal conductivity can be obtained at different temperatures and doping concentrations. The Boltztrap program has an input interface for VASP files, which can meet the needs of present high-throughput processes. After completing static calculations, the Boltztrap module can be performed. Meanwhile, Boltztrap based on Python can be well embedded into our high-throughput Python data processing scripts, which are written for quickly obtaining the calculated quantities such as Seebeck coefficient, electronic conductivity, and electronic thermal conductivity. Combined with the lattice thermal conductivities estimated from the elastic properties calculations, we could obtain the ZT values for the materials. We listed the top ten semiconductor materials with ZT values in Table 3.

**Table 3.** Top 10 semiconductor materials sorted by ZT value.

| id | Element | $\kappa_l(WK^{-1}m^{-1})$ | $S(\mu V/K)$ | $\sigma(kS/m)$ | $\kappa_e(WK^{-1}m^{-1})$ | ZT | Type |
|---|---|---|---|---|---|---|---|
| 10653 | ['Sr', 'Te'] | 2.0851 | 787.901 | 92496.65 | 1235.61 | 13.917 | p |
| 28110 | ['Rb', 'Pt', 'I'] | 0.2467 | 685.479 | 417.93 | 5.12 | 10.972 | n |
| 9319 | ['Ba', 'Pr', 'Pt', 'O'] | 5.3682 | 440.822 | 2528.83 | 8.06 | 10.970 | p |
| 30055 | ['Rb', 'Br', 'O'] | 2.0237 | 523.638 | 17534.51 | 149.09 | 9.544 | n |
| 28651 | ['Cs', 'Ir', 'Cl'] | 0.2638 | 378.607 | 115.54 | 0.30 | 8.702 | n |
| 14017 | ['K', 'Sb'] | 0.8422 | 411.902 | 1559.21 | 10.39 | 7.065 | n |
| 168 | ['Sn', 'Se'] | 0.0451 | 492.515 | 12.98 | 0.12 | 5.581 | p |
| 3060 | ['Cs', 'Pt', 'I'] | 0.2562 | 958.567 | 6.39 | 0.07 | 5.410 | n |
| 4783 | ['Ba', 'Pr', 'O'] | 10.8090 | 402.637 | 2925.30 | 17.09 | 5.097 | p |
| 30373 | ['Rb', 'Au'] | 1.9033 | 557.953 | 354.34 | 4.92 | 4.843 | p |

*3.6. ZT value and BE value*

As an example for the application of our database, we associate thermoelectric ZT values with the electronic quality factor. By $S$ and $\sigma$, the electronic quality factor $B_E$ can be defined by [22]:

$$B_E = S^2\sigma\left[\frac{S_r^2 exp(2-S_r)}{1+exp[5-5S_r]} + \frac{S_r\pi^2/3}{1+exp[5(S_r-1)]}\right] \quad (21)$$

where $S_r = |S|e/k_B$. As shown in Figure 6, the $ZT$ values of most materials are positively correlated to its electronic quality factor $B_ET/\kappa_L$, so the $B_ET/\kappa_L$ values can also serve as another criterion for judging excellent thermoelectric materials.

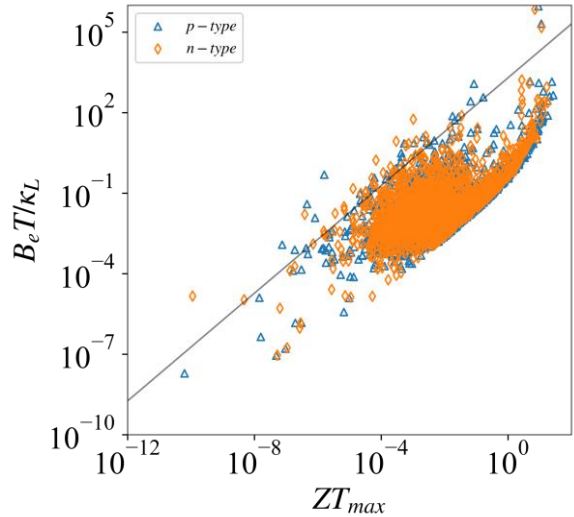

**Figure 6.** Thermal power quality factor $B_E T/\kappa_L$ and maximum $ZT_{max}$ at 300K.

## 4. Conclusions

In this work, we builds a thermoelectric material database——Wenzhou TE. We designed several modules to obtain the electronic and heat transport parameters for materials, including structural screening, deformation potential, elastic constant, and Boltztrap electrical transport performance calculations module. And we write several Python scripts to collect data and process results. Furthermore, we built a webpage for the first-principles calculated thermoelectric materials database (https://hezhu2024.github.io), which could be used for searching and viewing the physical properties of materials. Subsequently, we will continue the construction of the database to include more materials, and based on this, one can easily use these data for data mining and thermoelectric material development.

## 5. Acknowledgments

This study was partly supported by National Natural Science Foundation of China (52272006), Zhejiang Provincial Natural Science Foundation of China (LY22A040001), S &T Innovation 2025 Major Special Program of Ningbo (2020Z054), and Wenzhou Municipal Natural Science Foundation (G20210016).